\RequirePackage{fix-cm}
\documentclass[twocolumn]{svjour3}          
\pdfoutput=1
\smartqed  
\usepackage{graphicx}
\begin{document}

\title{Temperature effect on the electronic structure of the polaronic excitations within the three-band p-d-Holstein model}



\author{I.A. Makarov         \and
        S.G. Ovchinnikov 
}


\institute{I. Makarov and S. Ovchinnikov\at
              Kirensky Institute of Physics, Federal Research Center KSC SB RAS, 660036 Krasnoyarsk, Russia \\
              \email{sgo@iph.krasn.ru}
}

\date{Received: date / Accepted: date}

\maketitle

\begin{abstract}
In this work we investigate temperature dependence of electronic structure of system with strong electronic correlations and strong electron-phonon interaction modeling cuprates in the frameworks of the three-band p-d-Holstein model by a polaronic version of the generalized tight binding (GTB) method. Within this approach the electronic structure is formed by polaronic quasiparticles constructed as excitations between initial and final polaronic multielectron states. Temperature effect is taken into account by occupation numbers of local excited polaronic states and variations in the magnitude of spin-spin correlation functions. Temperature increasing leads to broadening of the spectral function peak at the top of the valence band, shift of the peak, the decreasing of the peak intensity.

\end{abstract}

\section{Introduction}
\label{sec_introduction}
There is still the unsolved question of how electronic structure of cuprates in the normal phase is formed from the undoped La$_2$CuO$_4$. Besides the strong electron correlation (SEC) that results in the Mott-Hubbard insulator ground state of La$_2$CuO$_4$, the undoped copper oxides may have also strong electron-phonon interaction (EPI). The experimental indications for the strong EPI have been found by the ARPES measurements in the compound Ca$_2$CuO$_2$Cl$_2$ that have revealed unusually large linewidth ($\sim1$ eV) in comparison to the narrow line in isostructural Sr$_2$RuO$_4$~\cite{ShenKM2004}. Width of the peak grows with temperature increasing, wherein its position is shifted deep into the band and its spectral weight is lowered~\cite{Kim2002,ShenKM2007}. Spectra in the optical experiments also demonstrates large temperature effect~\cite{Onose1999}. Explanation of the origin of such experimental features can serve as a key to understanding the nature of quasiparticles forming the electronic structure of cuprates.

Within several theoretical approaches it was confirmed that the key factor for description of ARPES spectra and its temperature dependence in the undoped cuprates is constructive interplay between strong EPI and strong Coulomb~\cite{Mishchenko2004,MishchenkoUFN,Rosch2005}. In Ref.~\cite{CatFilMisNag} a growth of the linewidth, binding energy enhancement and spectral weight damping with temperature increasing were obtained within the $t-J$-Holstein model in the frameworks of Hybrid Dynamical Momentum Average self-consistent method. Study of the $t-J$-Holstein model within adiabatic approximation~\cite{RoschGunEurPhysJ} in Ref.~\cite{Rosch2005} also results in the temperature dependence of linewidth.

The $t-J$ model is known to be the effective low-energy model for cuprates~\cite{Bulaevskii1968,Chao1977,Hirsch1985}. The multiphonon excitations in the systems with strong EPI results in the electron states beyond the low-energy window. That is why more general approach in the framework of the multiband $p-d$ model with SEC and strong EPI is desirable. In this work we will investigate electronic structure of polaronic excitations in the system with SEC and strong EPI the prototype of which is the undoped single-layer cuprate La$_2$CuO$_4$ in wide interval of temperatures (from $10$ to $760$~K). We will use three-band $p-d$-Holstein model within the polaronic Generalized Tight-Binding (p-GTB) method~\cite{Makarov2015}.

\section{Three-band $p-d$-Holstein model within polaronic GTB-method at finite temperature}
\label{Model_and_GTB}
The GTB method~\cite{Ovchinnikov89,Gav2000,LDA+GTB} is a cluster perturbation theory approach, it includes exact diagonalization of the Hamiltonian for a lattice of separate clusters with different number of fermions in each cluster, construction of Fermi-type excitations between multielectron eigenstates of the cluster described by Hubbard operators and perturbation treatment of all intercluster hoppings and interactions to obtain band structure. In the p-GTB method basis cluster states contain electron and phonon wave functions, eigenstates are polaronic states and excitations between local eigenstates are multiphonon Franck-Condon processes~\cite{Makarov2015}.

It is widely accepted that a low-energy electronic structure of HTSC cuprates is formed by distribution of holes in the CuO$_2$ plane over copper ${d_{{x^2} - {y^2}}}$ (hereinafter $d$) and oxygen ${p_{x,y}}$ orbitals. Phonon system will be described by dispersionless local vibrations, light O atoms vibrate relative to the fixed heavy Cu atoms. EPI is written as a renormalization of on-site energy on Cu atom caused by oxygen atom displacement. Thus the three-band $p-d$-Holstein model is written as:
\begin{eqnarray}
\label{pd_Hamiltonian}
H & = & {H_{el}} + {H_{ph}} + {H_{e - ph}} \nonumber \\
{H_{el}} & = & \sum\limits_{\bf{f}\sigma } {{\varepsilon _d}d_{\bf{f}\sigma }^\dag {d_{\bf{f}\sigma }}}  + \sum\limits_{\alpha \bf{h}\sigma } {{\varepsilon _p}p_{\alpha\bf{h}\sigma }^\dag {p_{\alpha \bf{h}\sigma }}}  + \nonumber \\
& & + \sum\limits_{\bf{fh}\alpha \sigma } {{{\left( { - 1} \right)}^{{R_{\bf{h}}}}}{t_{pd}}\left( {d_{\bf{f}\sigma }^ \dag {p_{\alpha\bf{h}\sigma }} + h.c.} \right)}  + \nonumber \\
& & + \sum\limits_{\alpha \alpha '\bf{h} \ne \bf{h'}\sigma } {{{\left( { - 1} \right)}^{{M_{\bf{hh'}}}}}{t_{pp}}\left( {p_{\alpha\bf{h}\sigma }^\dag {p_{{\alpha '}\bf{h'}\sigma }} + h.c.} \right)}  + \nonumber \\
& & + \sum\limits_{\bf{f}} {{U_d}d_{\bf{f} \uparrow }^\dag {d_{\bf{f} \uparrow }}d_{\bf{f} \downarrow }^\dag {d_{\bf{f} \downarrow }}}  + \nonumber \\
& & + \sum\limits_{\bf{h}} {{U_p}p_{\alpha\bf{h} \uparrow }^\dag {p_{\alpha\bf{h} \uparrow }}p_{\alpha\bf{h} \downarrow }^\dag {p_{\alpha\bf{h} \downarrow }}}  + \nonumber \\
& & + \sum\limits_{\alpha\bf{fh}\sigma \sigma '} {{V_{pd}}d_{\bf{f}\sigma }^\dag {d_{\bf{f}\sigma }}p_{\alpha\bf{h}\sigma '}^\dag {p_{\alpha\bf{h}\sigma '}}} \nonumber \\
{H_{ph}} & = &\frac{M_O}{2}\sum\limits_{\bf{h}} {\left( {\dot u_{\bf{h}}^2 + {\omega_O}^2\dot u_{\bf{h}}^2} \right)} \nonumber \\
{H_{e-ph}} & = & \sum\limits_{\bf{f}\sigma } {{\left( {\sum\limits_{\bf{h}} {{{\left( { - 1} \right)}^{{S_{\bf{h}}}}}{g_d}{u_{\bf{h}}}} } \right)} d_{\bf{f}\sigma }^\dag {d_{\bf{f}\sigma }}}
\end{eqnarray}
Here ${d_{\bf{f}\sigma }}$ and ${p_{\alpha\bf{h}\sigma }}$ are the operators of hole annihilation with spin $\sigma $ on $d$-orbital of the copper atom ${\bf{f}}$ and ${p_x}$(${p_y}$)-orbital of the oxygen atom ${\bf{h}}$, respectively. ${\bf{h}}$ runs over two of the four positions of planar oxygen atoms neighboring to Cu atom in octahedral unit cell centered on site ${\bf{f}}$ at each $\alpha$, ${\bf{h}} = {\left( {{f_x} \pm {a \mathord{\left/
 {\vphantom {a 2}} \right.
 \kern-\nulldelimiterspace} 2}, {f_y}} \right)}$ if $\alpha=x$ and ${\bf{h}} = {\left( {{f_x},{f_y} \pm {b \mathord{\left/
 {\vphantom {b 2}} \right.
 \kern-\nulldelimiterspace} 2}} \right)}$ if $\alpha=y$, $a$ and $b$ are the lattice parameters. ${\varepsilon _d}$ is the on-site energy of hole on Cu ion and ${\varepsilon _p}$ is the same on O ion; $t_{pd}$ is the amplitude of nearest-neighbor hopping between $d$-orbitals of Cu ion ${\bf{f}}$ and ${p_{x,y}}$-orbitals of O ion ${\bf{h}}$ in CuO$_2$ plane and $t_{pp}$ is the amplitude of nearest-neighbor hopping between ${p_{x,y}}$-orbitals of the oxygen atoms ${\bf{h}}$ and ${\bf{h'}}$. The phase parameters ${R_{\bf{h}}}$ and ${M_{\bf{hh'}} }$ are determined by phases of overlapping wave functions. ${U_d}$ is the Coulomb interaction of two holes on the same copper atom and ${U_p}$ is the same for oxygen atom, $V_{pd}$ is the intersite Coulomb interaction. We will use the following parameters of electronic Hamiltonian ${H_{el}}$ (in eV):
\begin{eqnarray}
{\varepsilon _d} = 0, {\varepsilon _p} = 1.5, {t_{pd}} = 1.36, {t_{pp}} = 0.86 \nonumber \\
{U_d} = 20, {U_p} = 18, {V_{pd}} = 14,
\label{parameters}
\end{eqnarray}
on-site energies and hopping integrals were obtained for La$_2$CuO$_4$ from LDA+GTB method~\cite{LDA+GTB}. In $H_{ph}$ and $H_{e-ph}$ ${u_{\bf{h}}} = \xi \left( {e_{\alpha\bf{h}}^ \dag  + {e_{\alpha\bf{h}}}} \right)$ is the operator of oxygen atom ${\bf{h}}$ displacement, where $\xi  = \sqrt {\frac{\hbar }{{2{M_O}{\omega_O}}}} $, ${M_O}$ is the mass of oxygen atom. $e_{\alpha\bf{h}}^\dag $ is the operator of creation of local phonon with frequency ${\omega_O}$, $\alpha$ denotes direction of atom ${\bf{h}}$ displacement. Phonon energy $\hbar {\omega_O} = 0.04$~eV. For oxygen atom ${\bf{h}} = \left( {{f_x} \pm {a \mathord{\left/
 {\vphantom {a 2}} \right.
 \kern-\nulldelimiterspace} 2},{f_y}} \right)$ (${\bf{h}} = \left( {{f_x},{f_y} \pm {b \mathord{\left/
 {\vphantom {b 2}} \right.
  \kern-\nulldelimiterspace} 2}} \right)$) displacement is along $\alpha=x$ ($y$) axis. ${g_d}$ is the parameter of the diagonal EPI between hole on $d$-orbital and phonon. The phase parameter ${S_{\bf{h}} } = 0$ for ${\bf{h}} = \left( {{f_x} + {a \mathord{\left/
 {\vphantom {a 2}} \right.
 \kern-\nulldelimiterspace} 2}, {f_y}} \right), \left( {{f_x}, {f_y} + {b \mathord{\left/
 {\vphantom {b 2}} \right.
 \kern-\nulldelimiterspace} 2}} \right)$ and ${S_{\bf{h}} } = 1$ for ${\bf{h}} = \left( {{f_x} - {a \mathord{\left/
 {\vphantom {a 2}} \right.
 \kern-\nulldelimiterspace} 2}, {f_y}} \right), \left( {{f_x}, {f_y} - {b \mathord{\left/
 {\vphantom {b 2}} \right.
 \kern-\nulldelimiterspace} 2}} \right)$, it is consistent with modulation of the on-site energy. We introduce dimensionless EPI parameter ${\lambda _{d}} = {{{{\left( {{g_{d}}\xi } \right)}^2}} \mathord{\left/
 {\vphantom {{{{\left( {{g_{d}}\xi } \right)}^2}} {W\hbar {\omega_O}}}} \right.
 \kern-\nulldelimiterspace} {W\hbar {\omega_O}}}$, $W$ is the bandwidth of the free electron in tight-binding method without EPI, we accept here $W = 2.6$~eV.

Since each planar oxygen atom belongs to the two CuO$_6$ clusters at once we should make orthogonalization procedure. We proceed from the hole $\left| {{p_{x{\bf{h}}}}} \right\rangle $, $\left| {{p_{y{\bf{h}}}}} \right\rangle $ and phonon $e_{x\bf{h}}^\dag \left| 0 \right\rangle $, $e_{y\bf{h}}^\dag \left| 0 \right\rangle $ atomic oxygen orbitals to the molecular oxygen orbitals $\left| {{b_{\bf{f}}}} \right\rangle $, $\left| {{a_{\bf{f}}}} \right\rangle $ and $A_{\bf{f}}^\dag \left| 0 \right\rangle $, $B_{\bf{f}}^\dag \left| 0 \right\rangle $ by transformation in the $k$-space~\cite{Shastry89}:
\begin{eqnarray}
\label{Shastry}
{b_{\bf{k}}}&=& \frac{i}{{{\mu _{\bf{k}}}}}\left( {{s_{{\bf{k}}x}}{p_{x{\bf{k}}}} - {s_{{\bf{k}}y}}{p_{y{\bf{k}}}}} \right)\nonumber \\
{a_{\bf{k}}}&=& - \frac{i}{{{\mu _{\bf{k}}}}}\left( {{s_{{\bf{k}}y}}{p_{x{\bf{k}}}} + {s_{{\bf{k}}x}}{p_{y{\bf{k}}}}} \right)\nonumber \\
{A_{\bf{k}}}&=&  - \frac{i}{{{\mu _{\bf{k}}}}}\left( {{s_{{\bf{k}}x}}{e_{x{\bf{k}}}} + {s_{{\bf{k}}y}}{e_{y{\bf{k}}}}} \right)\nonumber \\
{B_{\bf{k}}}&=& - \frac{i}{{{\mu _{\bf{k}}}}}\left( {{s_{{\bf{k}}y}}{e_{x{\bf{k}}}} - {s_{{\bf{k}}x}}{e_{y{\bf{k}}}}} \right)
\end{eqnarray}
where ${s_{{\bf{k}}x}} = \sin \left( {{{{k_x}a} \mathord{\left/
 {\vphantom {{{k_x}a} 2}} \right.
 \kern-\nulldelimiterspace} 2}} \right)$, ${s_{{\bf{k}}y}} = \sin \left( {{{{k_y}b} \mathord{\left/
 {\vphantom {{{k_y}b} 2}} \right.
 \kern-\nulldelimiterspace} 2}} \right)$ and ${\mu _{\bf{k}}} = \sqrt {s_{{\bf{k}}x}^2 + s_{{\bf{k}}y}^2} $.

After orthogonalization procedure Hamiltonian $H$ can be divided on the intracluster part and intercluster interactions:
\begin{equation}
\label{Hc_Hcc_tot}
H = {H_c} + {H_{cc}}, {H_c} = \sum\limits_{\bf{f}} {{H_{\bf{f}}}}, {H_{cc}} = \sum\limits_{\bf{fg}} {{H_{\bf{fg}}}}
\end{equation}

\begin{figure}
\center
\includegraphics[width=0.9\linewidth]{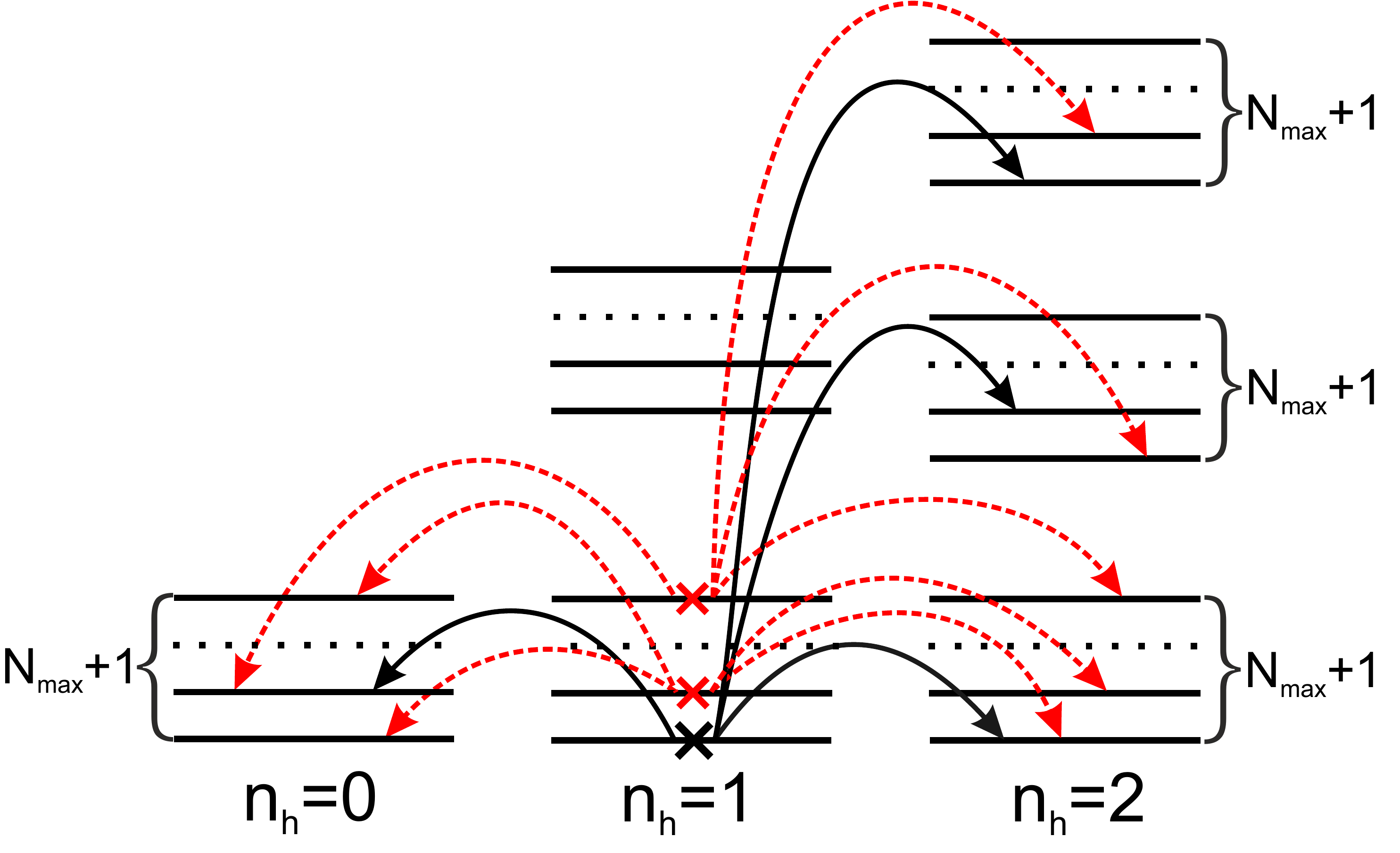}
\caption{\label{fig:levels} Schematic picture of the multielectron vibronic eigenstates of CuO$_6$ cluster (black horizontal lines) with hole numbers ${n_h} = 0,1,2$ and Fermi-type excitations between them. Black solid arrows denotes excitations with nonzero spectral weight at $T = 0$ K due to occupation of single-hole ground state (black cross) for undoped La$_2$CuO$_4$. At nonzero temperature excited single-hole states are thermally occupied (red crosses) and excitations involving these states acquire spectral weight (red dashed arrows).}
\end{figure}

Eigenstates of the CuO$_6$ cluster with hole numbers ${n_h} = 0,1,2$ obtained by exact diagonalization of Hamiltonian ${H_{\bf{f}}}$ include hole and phonon basis wave functions:
\begin{eqnarray}
|0,\nu \rangle & = & |0\rangle |\nu \rangle, \nu  = 0,1,...,{N_{\max }},\nonumber \\
\left| {1\sigma ,i} \right\rangle & = & \sum\limits_{\nu  = 0}^{N_{max}} {\left( {c_{i\nu }^d\left| {{d_\sigma }} \right\rangle \left| \nu  \right\rangle  + c_{i\nu}^b\left| {{b_\sigma }} \right\rangle \left| \nu  \right\rangle } \right)},\nonumber \\
\left| {2,j} \right\rangle & = & \sum\limits_{\nu  = 0}^{N_{max}} \left(  c_{j\nu }^{ZR}\left| {{\rm{ZR}}} \right\rangle \left| \nu  \right\rangle + \right. \nonumber \\
 &+ & \left. c_{j\nu }^{dd}\left| {{d_{\downarrow }}{d_{\uparrow }}} \right\rangle \left| \nu  \right\rangle  + c_{j\nu }^{bb}\left| {{b_ \downarrow }{b_ \uparrow }} \right\rangle \left| \nu  \right\rangle \right)
\label{es}
\end{eqnarray}
Indexes $i,j$ numerate the ground and excited eigenstates, {\rm{ZR}} is the Zhang-Rice singlet state, $|\nu \rangle $ - phonon state with phonon number ${n_{ph}} = \nu $. Phonon state $\left| \nu  \right\rangle $ are $\nu $-times action of phonon creation operator ${A^\dag }$ on vacuum state of harmonic oscillator:
\begin{equation}
\left| \nu  \right\rangle  = \frac{1}{{\sqrt {\nu !} }}{\left( {{A^\dag }} \right)^\nu }\left| {0,0,...,0} \right\rangle
\label{harm_osc}
\end{equation}
 ${N_{\max }}$ is the cutoff for number of phonons, it is calculated for each certain set of parameters from the following condition: addition of phonon number above ${N_{\max }}$, $N > {N_{\max }}$, does not change the electron spectral function. The value ${N_{\max }}$ mainly depends on the EPI coupling parameter. Similar description of the local polaronic states has been discussed previously~\cite{Piekarz1999}.

A scheme of multielectron and multiphonon CuO$_6$ cluster levels~(\ref{es}) is depicted in Fig.~\ref{fig:levels}. The Fermi-type multiphonon excitations (arrows in Fig.~\ref{fig:levels}) describe the Franck-Condon processes. Each excitation from cluster eigenstate $\left| q \right\rangle $ to final eigenstate $\left| p \right\rangle $ can be described by the Hubbard operator $X_{\bf{f}}^{pq} = \left| p \right\rangle \left\langle q \right|$. The Fermi-type operators of annihilation of hole on copper and oxygen orbital can be expressed in terms of the Fermi-type Hubbard operators $X_{\bf{f}}^{pq}$:
\begin{eqnarray}
{d_{{\bf{f}}\sigma }} = \sum\limits_{pq} {{\gamma _{d\sigma }}\left( {pq} \right)} X_{\bf{f}}^{pq} \nonumber \\
{b_{{\bf{f}}\sigma }} = \sum\limits_{pq} {{\gamma _{b\sigma }}\left( {pq} \right)} X_{\bf{f}}^{pq}
\label{Xop}
\end{eqnarray}
The phonon annihilation operator is expressed through the Bose-type Hubbard operators $Z_{\bf{f}}^{pp'}$:
\begin{equation}
{A_{\bf{f}}} = \sum\limits_{pp'} {{\gamma _A}\left( {pp'} \right)} Z_{\bf{f}}^{pp'}
\label{Zop}
\end{equation}
where states $\left| p \right\rangle $ and $\left| {p'} \right\rangle $ have the same number of holes and belong to the same sector of Hilbert space $\left| {{n_h}} \right\rangle $. Since we will study the electronic states close to the top of the valence band that in our model is the electronic lower Hubbard band (LHB), the Franck-Condon processes will be denoted by ``index of single-hole eigenstate - index of two-hole eigenstate'', $i - j$.

Hamiltonians ${H_c}$ and ${H_{cc}}$ in Eq.~(\ref{Hc_Hcc_tot}) are rewritten in the terms of Hubbard operators:
\begin{eqnarray}
& {H_c} & = \sum\limits_{\bf{f}} {\left[ {\sum\limits_l {{\varepsilon _{0l}}Z_{\bf{f}}^{0l,0l}}  + \sum\limits_i {{\varepsilon _{1i}}Z_{\bf{f}}^{1i,1i}}  + \sum\limits_j {{\varepsilon _{2j}}Z_{\bf{f}}^{2j,2j}} } \right]} \nonumber \\
& {H_{cc}} & = \sum\limits_{{\bf{f}} \ne {\bf{g}}} {\left[ {\sum\limits_{mn} {2{t_{pd}}{\mu _{{\bf{fg}}}}\gamma _{{d_x}}^ * \left( m \right){\gamma _b}\left( n \right)\mathop {X_{\bf{f}}^m}\limits^\dag  X_{\bf{g}}^n} } \right.}  - \nonumber \\
& &\left. { - \sum\limits_{mn} {2{t_{pp}}{\nu _{{\bf{fg}}}}\gamma _b^ * \left( m \right){\gamma _b}\left( n \right)\mathop {X_{\bf{f}}^m}\limits^\dag  X_{\bf{g}}^n} } \right]
\label{Hc_Hcc_Xop}
\end{eqnarray}
Here ${\varepsilon _{0l}}$, ${\varepsilon _{1i}}$, ${\varepsilon _{2j}}$ are the energies of cluster eigenstates with ${n_h} = 0,1,2$. The intercluster interactions result from the $p-d$ and $p-p$ hoppings of the Hubbard polarons between clusters, we consider hopping up to sixth neighbors. To obtain the band dispersion and spectral function of Hubbard polarons we use the equation of motion for the Green function ${D^{mn}}\left( {{\bf{f}},{\bf{g}}} \right) = \left\langle {\left\langle {{X_{\bf{f}}^m}}
 \mathrel{\left | {\vphantom {{X_{\bf{f}}^m} {X_{\bf{g}}^n}}}
 \right. \kern-\nulldelimiterspace}
\mathop {X_{\bf{g}}^n}\limits^\dag \right\rangle } \right\rangle $, where $m$,$n$ are the quasiparticle band indexes, this index is uniquely defined by initial and final states of excitation $m \equiv \left( {p,q} \right)$. The set of equations of motion is decoupled in the generalized Hartri-Fock approximation by method of irreducible Green functions~\cite{Plakida1989,Yushankhay1991,Valkov2005,KorshOvch2007} taking into account the interatomic spin correlation functions. The Dyson equation for the matrix Green function $\hat D\left( {{\bf{f}},{\bf{g}}} \right)$ in the momentum space has the form
\begin{equation}
\hat D\left( {{\bf{k}};\omega } \right) = {\left[ {\hat D_0^{ - 1}\left( \omega  \right) - \hat F\hat {\tilde t}\left( {{\bf{k}}} \right) + \hat \Sigma \left( {{\bf{k}};\omega } \right)} \right]^{ - 1}}\hat F
\label{Dyson_eq}
\end{equation}
In this equation ${\hat D_0}$ is the exact local Green function, its matrix elements $D_0^{mn} = {{{\delta _{mn}}F\left( m \right)} \mathord{\left/
 {\vphantom {{{\delta _{mn}}F\left( m \right)} {\left( {\omega  - \Omega \left( m \right)} \right)}}} \right.
 \kern-\nulldelimiterspace} {\left( {\omega  - \Omega \left( m \right)} \right)}}$, $\Omega \left( m \right) = \Omega \left( {pq} \right) = {\varepsilon_p} - {\varepsilon_q}$ is the energy of the quasiparticle excitation between states $q$ and $p$, matrix elements $F^{mn} = F\left( m \right){\delta _{mn}}$, $F\left( m \right) = F\left( {pq} \right) = \left\langle {{Z^{pp}}} \right\rangle  + \left\langle {{Z^{qq}}} \right\rangle $ is the filling factor of the quasiparticle. The matrix of intercluster hopping $\hat {\tilde t}\left( {{\bf{k}};\omega } \right)$ is determined by $p-d$ and $p-p$ hoppings, $\tilde t_{\bf{k}}^{mn} = \sum\limits_{\lambda \lambda '} {\gamma _\lambda ^ * \left( m \right)} {\gamma _{\lambda '}}\left( n \right)\tilde t_{\bf{k}}^{\lambda \lambda '}$. $\hat \Sigma \left( {{\bf{k}};\omega } \right)$ is the self-energy operator containing spin correlation functions.

Averages $\left\langle {{Z^{pp}}} \right\rangle$ is self-consistently determined from the condition of completeness $\sum\limits_{{n_h} p} {Z_{\bf{f}}^{{n_h}p,{n_h}p} = 1} $ and the chemical potential equation (here $n$ is the hole concentration for La$_{2-x}$Sr$_x$CuO$_4$):
\begin{eqnarray}
n = 1 + x = \sum\limits_{{n_h} p} {{n_h} \cdot \left\langle {{Z^{{n_h}p,{n_h}p}}} \right\rangle }
\label{chempot_eq}
\end{eqnarray}
where $p$ runs over all eigenstates in the corresponding Hilbert space sector and from the
Boltzmann distribution ${n_{1i}} = {n_{10}}\exp \left( { - \frac{{{\varepsilon_{1i}} - {\varepsilon_{10}}}}{{kT}}} \right)$ for a given temperature. At $T=0$~K only the ground single-hole state is filled and hence only excitations involving the ground single-hole state $\left| {1,0} \right\rangle $ have a non-zero spectral weight (solid black arrows in Fig.~\ref{fig:levels}). Filling of ground state falls whereas occupation of $1^{th}$, $2^{th}$ and $3^{rd}$ excited single-hole states monotonically grows with increasing temperature.

We will describe the short-range order in the paramagnetic state as the isotropic spin liquid with zero spin projections and non-zero spin correlation functions following to Refs.~\cite{Valkov2005,Shimahara1991,Barabanov1994}. Spin correlators for doping $x=0.01$ at $T=0$~K was taken from Ref.~\cite{KorshOvch2007} and their damping with temperature increasing is taken from Ref.~\cite{Shimahara1991}.

\section{Temperature dependence of the electronic structure in the system without EPI \label{T_dep_without_EPI}}
\begin{figure}
\center
\includegraphics[width=0.9\linewidth]{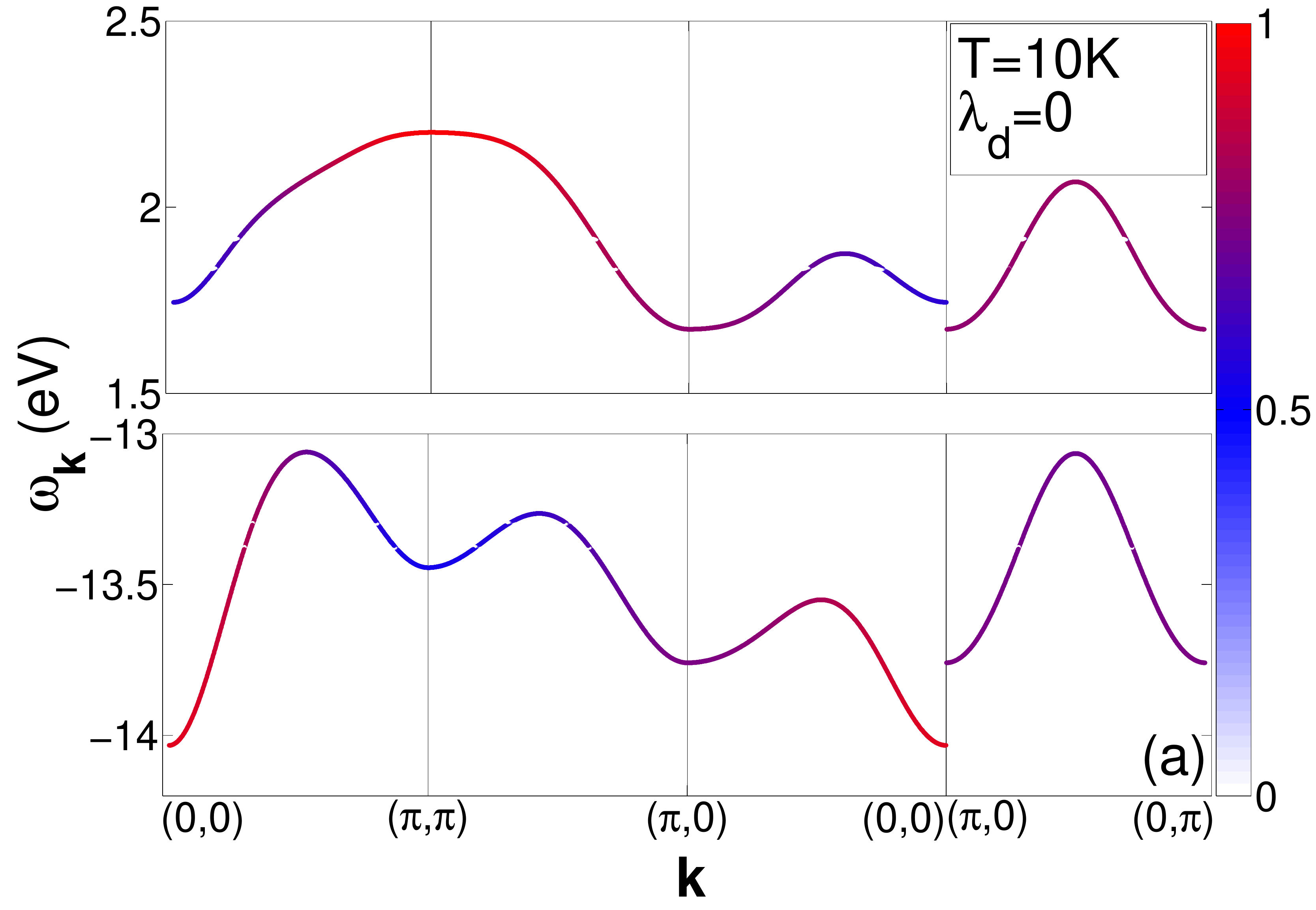}
\includegraphics[width=0.9\linewidth]{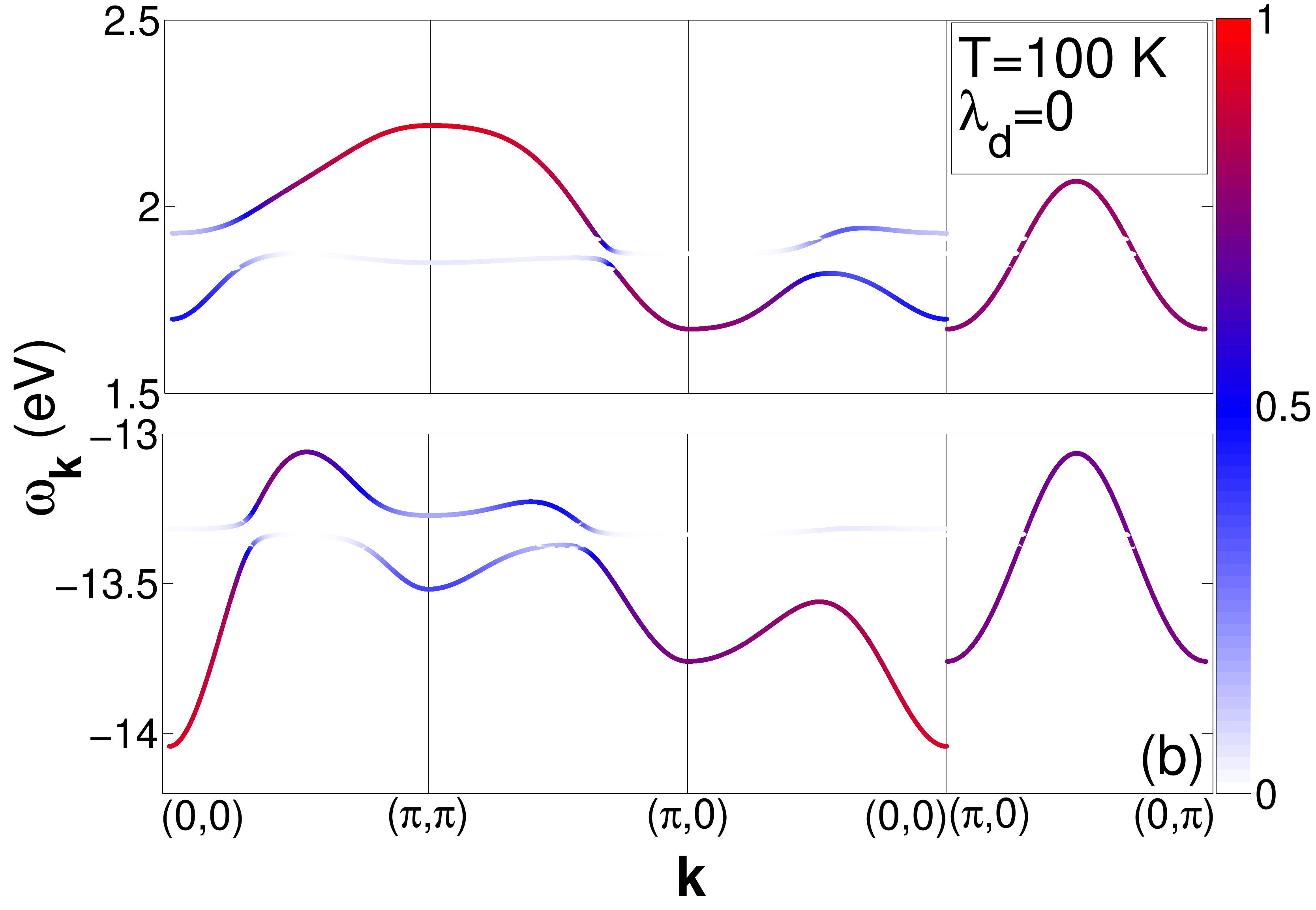}
\includegraphics[width=0.9\linewidth]{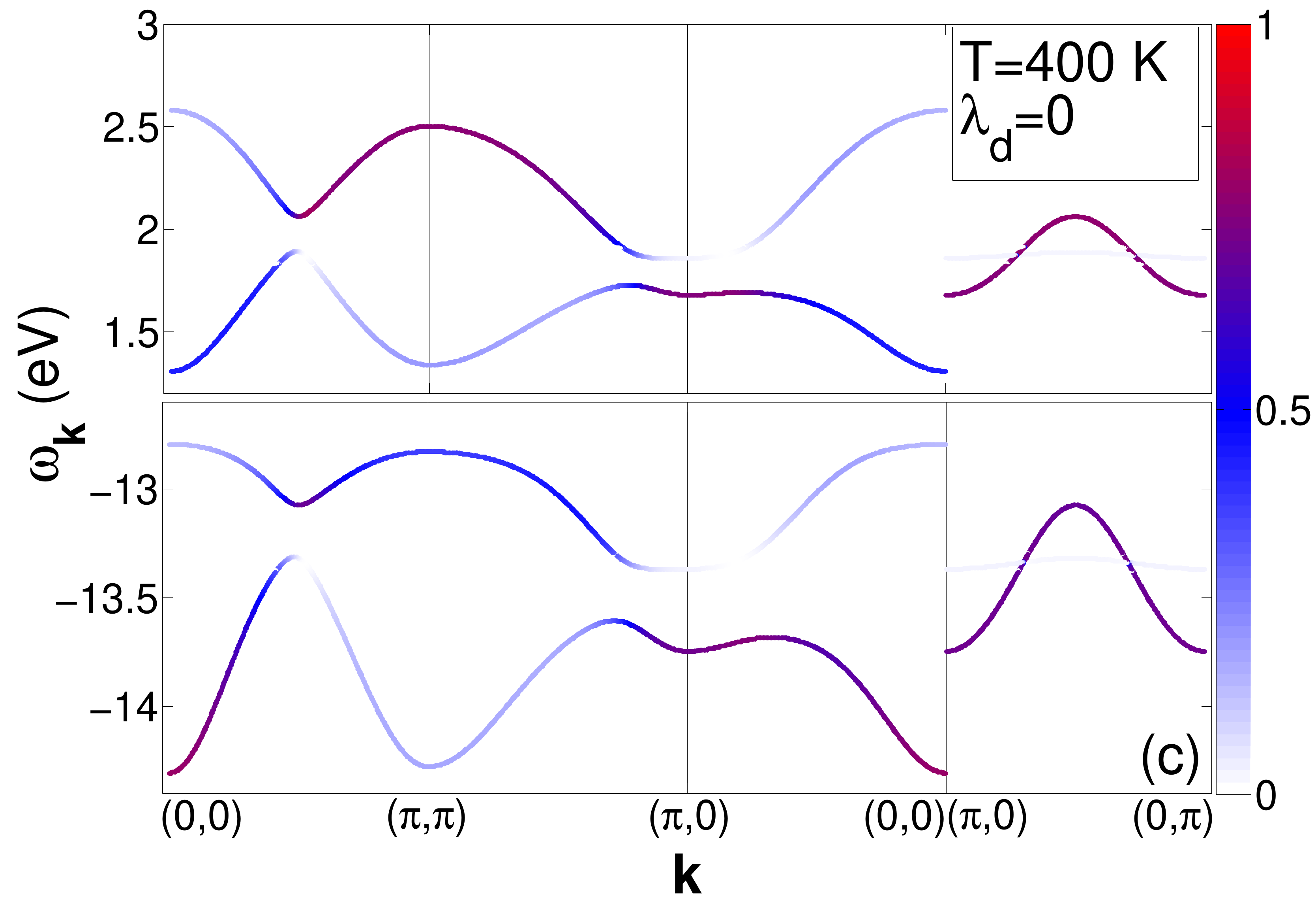}
\caption{\label{fig:bandstr_Tdep_withoutEPI} Evolution of the band structure of the quasiparticle excitations with increasing temperature for the system without EPI (${\lambda _d} = 0$). (a) $T = 10$~K, (b) $T = 100$~K, (c) $T = 400$~K. The conductivity (valence) bands are shown on the upper (lower) panel of each figure. Color in each $k$-point indicates the spectral weight of quasiparticle.}
\end{figure}

\begin{figure*}
\center
\includegraphics[width=0.45\linewidth]{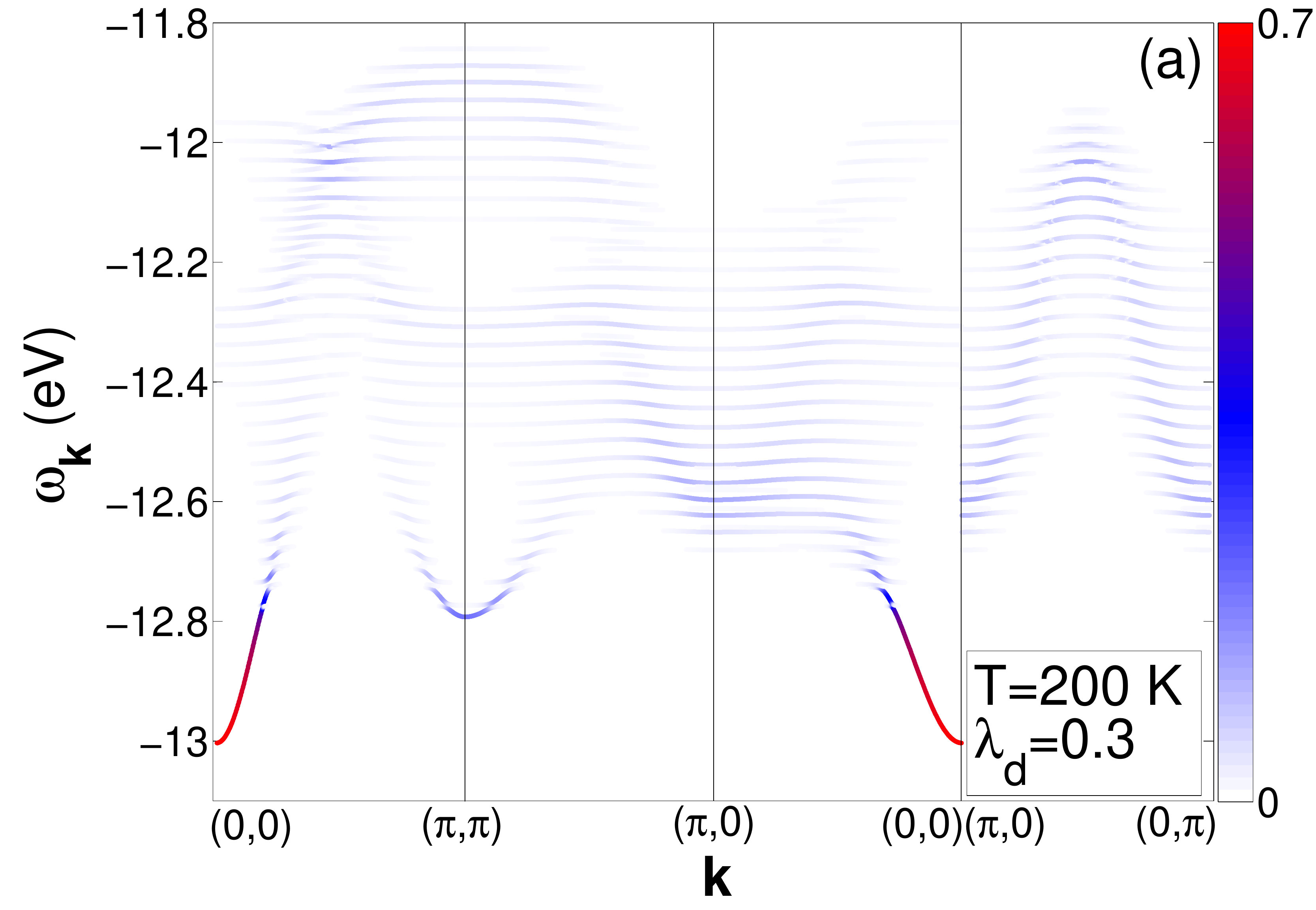}
\includegraphics[width=0.45\linewidth]{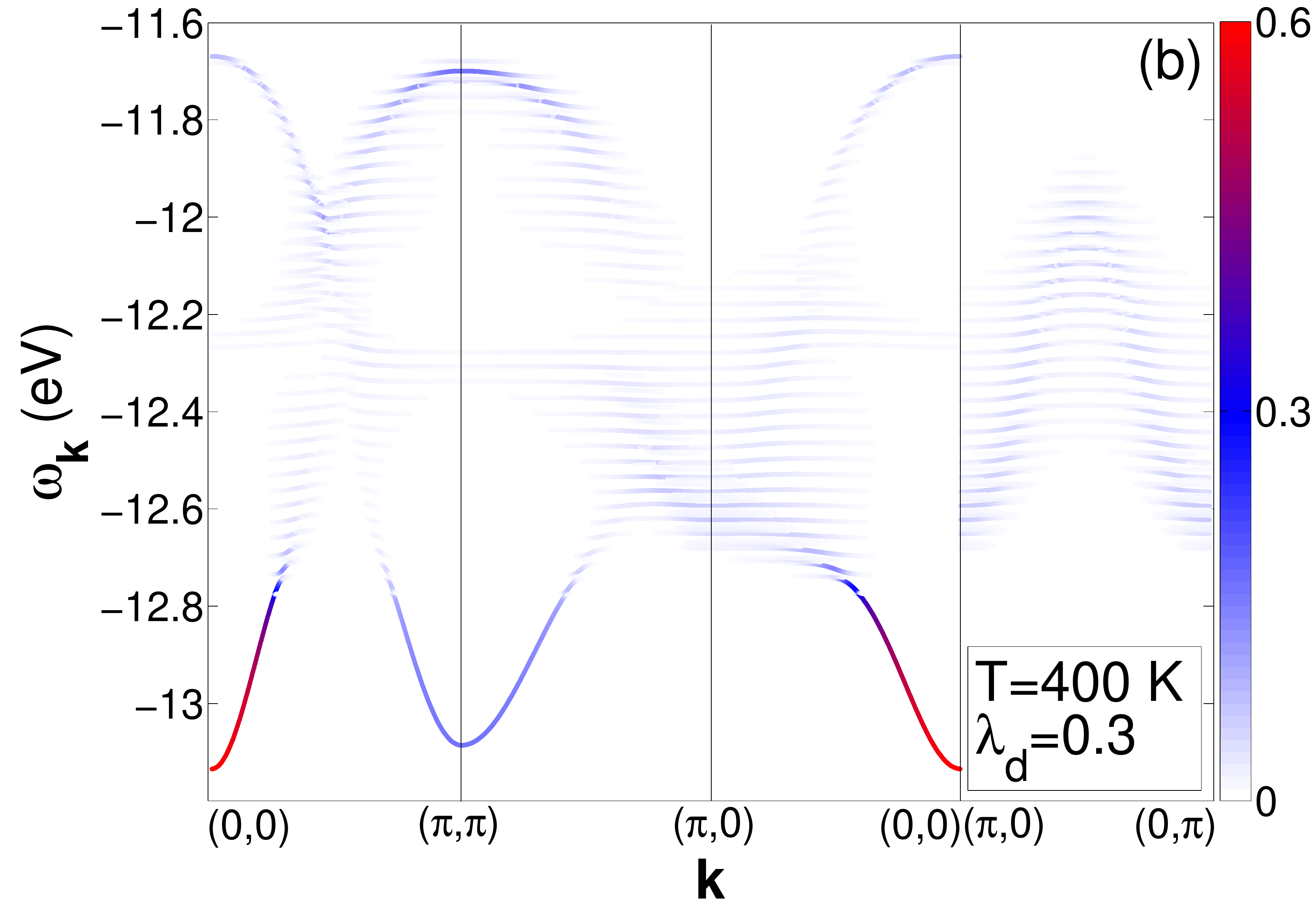}
\includegraphics[width=0.45\linewidth]{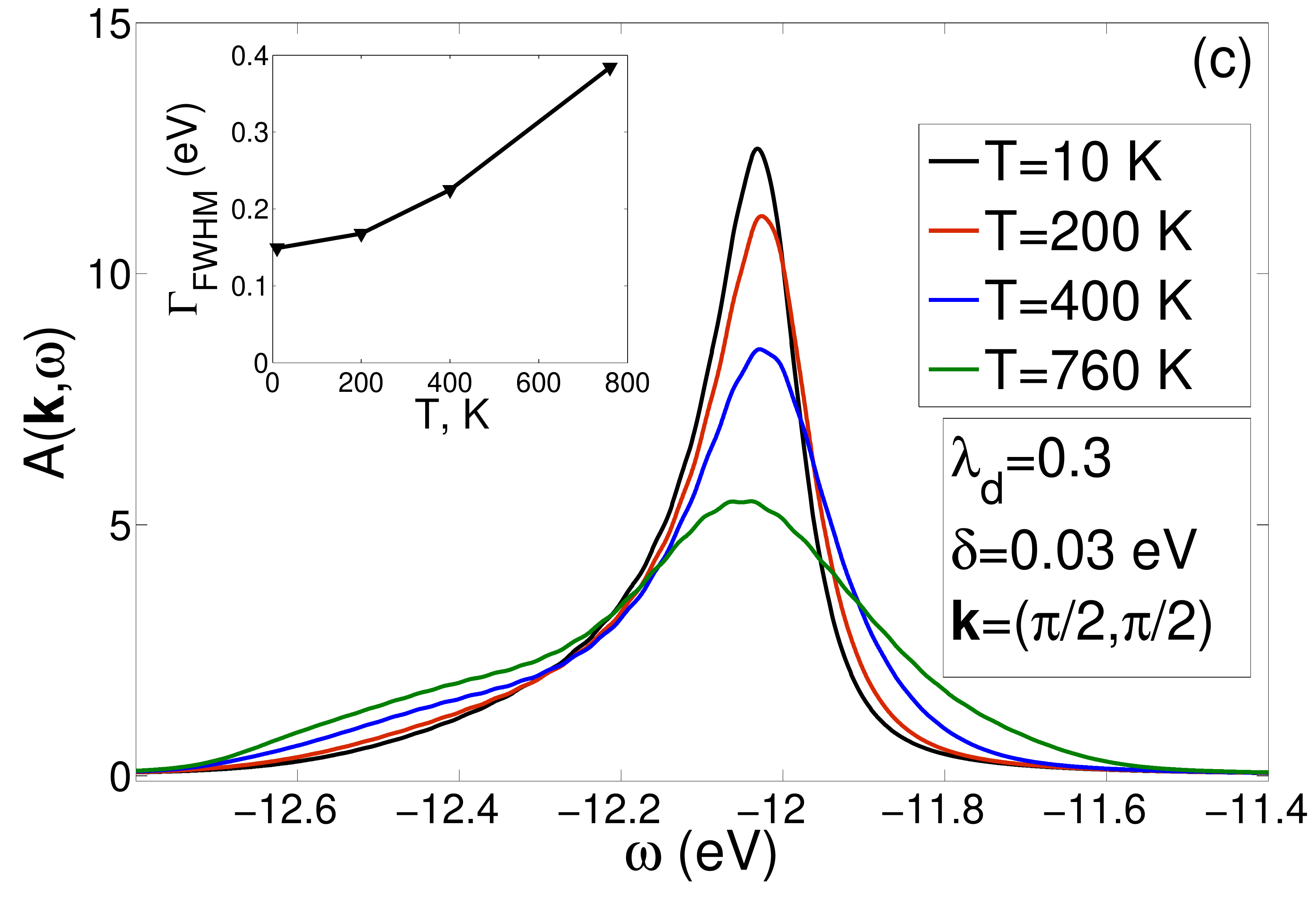}
\includegraphics[width=0.45\linewidth]{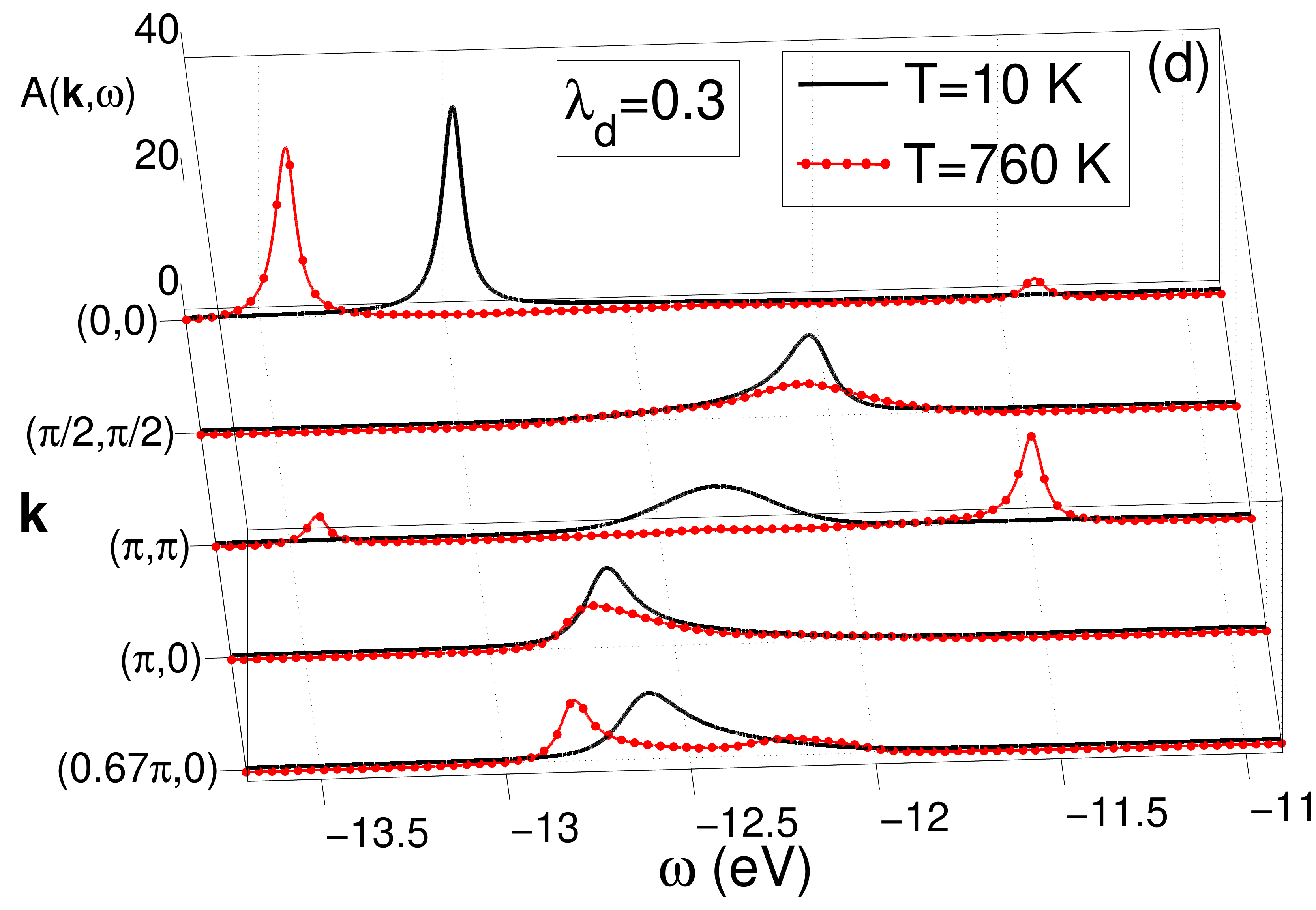}
\caption{\label{fig:elstr_Tdep_withEPI} Evolution of electronic structure at EPI constant ${\lambda _d} = 0.3$ with increasing temperature. Band structure of the valence band with spectral weight of quasiparticles in each $k$-point (spectral weight is displayed by color) at (a) $T = 200$~K, (b) $T = 400$~K. Note that spectral weight scale decreases with increasing temperature. (c) Broadening of spectral function peak (inset), damping of the its intensity and shift of the peak with increasing temperature are shown at ${\bf{k}} = \left( {\frac{\pi }{2},\frac{\pi }{2}} \right)$ for different temperatures ($T = 10$~K, $T = 200$~K, $T = 400$~K, $T = 760$~K). (d) Polaron spectral functions at $T = 10$~K and $T = 760$~K for ${\lambda _d} = 0.3$ at different k-points. Lorentzian width $\delta  = 0.03$~eV.}
\end{figure*}

Without EPI the LHB band structure at $T=0$ is formed by $0-0$ quasiparticle (excitation between single-hole and two-hole ground states without phonons) band and set of dispersionless degenerate Franck-Condon resonances with zero spectral weight. The thermal occupation of the excited single-hole states and non-zero matrix elements due to the interband hopping ${\tilde t^{mn}}\left( {m \ne n} \right)$ will result in the hybridization of the Hubbard band $0-0$ and the phononless Franck-Condon excitations $1-1$, $2-2$, $3-3$ etc. (Fig.~\ref{fig:bandstr_Tdep_withoutEPI}). Spectral weight of these pure electronic phononless excitations grows and is redistributed between them with increasing temperature. As a result of quasiparticles hybridization the $0-0$ band is splitted at the energy $\Omega \left( {1 - 1} \right)$ on two parts (Fig.~\ref{fig:bandstr_Tdep_withoutEPI}(b)). The upper part of the valence band is strongly modified under temperature increasing. Near $T=200$~K the local maximum at the point ${\bf{k}} = \left( {{\pi  \mathord{\left/
 {\vphantom {\pi  2}} \right.
 \kern-\nulldelimiterspace} 2},{\pi  \mathord{\left/
 {\vphantom {\pi  2}} \right.
 \kern-\nulldelimiterspace} 2}} \right)$ becomes local minimum by energy increase at points $\Gamma  = \left( {0,0} \right)$ and ${\rm M} = \left( {\pi ,\pi } \right)$. Further temperature increase results in energy growth at $\Gamma $ and ${\rm M}$ points (Fig.~\ref{fig:bandstr_Tdep_withoutEPI}(c)). The width of the whole band is increased.

\section{Temperature dependence of the electronic structure of the polaron quasiparticles \label{T_dep_with_EPI}}

The EPI leads to nonzero probability of multiphonon excitations ($0-1$, $0-2$, $0-3$ etc.). These excitations acquire spectral weight, weak dispersion and begin to interact with $0-0$ polaron. Therefore a number of Hubbard polaron subbands in the valence band (and in the conductivity band) appears. Electron spectral weight transfers from the coherent dispersive high-intensity $0-0$ quasiparticle to the incoherent multiphonon excitations with the EPI coupling increasing. Electron spectral function
\begin{eqnarray}
A\left( {{\bf{k}},\omega } \right) & = &\left( { - \frac{1}{\pi }} \right) \sum \limits_{\lambda \lambda '\sigma mn} {\gamma _{\lambda '\sigma }^ * \left( n \right){\gamma _{\lambda \sigma }}\left( m \right) \times } \nonumber \\
& \times & {\mathop{\rm Im}\nolimits} {\left\langle {\left\langle {{X_{\bf{k}}^m}}
 \mathrel{\left | {\vphantom {{X_{\bf{k}}^m} {\mathop {X_{\bf{k}}^n}\limits^\dag  }}}
 \right. \kern-\nulldelimiterspace}
 {{\mathop {X_{\bf{k}}^n}\limits^\dag  }} \right\rangle } \right\rangle _{\omega  + i\delta }}
\label{SpectralFunction}
\end{eqnarray}
at each $k$-point is formed by several peaks reflecting the Franck-Condon excitations. Taking into account finite lifetime of quasiparticles the multipeak structure of spectral function can be transformed into single broad peak if the EPI constant is large enough. In spite of large number of the Franck-Condon excitations crossing and lost of the coherency the LHB original dispersion of electronic model without EPI is mainly preserved.

At zero temperature the main peak of spectral function is formed by excitation from the occupied ground single-hole eigenstate to excited multiphonon polaron two-hole state for which the Franck-Condon factor is largest. Filling of excited single-hole states with one, two, three etc. thermal phonons grows with increasing temperature and such excitations as $1-0$, $2-0$, $1-2$ etc. acquire spectral weight. Spectral weight of the quasiparticles which are primary formed by these new multiphonon excitations involving excited eigenstates also increases. Generally contributions of different multiphonon excitations to complex quasiparticle forming specific polaron subband are redistributed with changing temperature. Peaks of emerged quasiparticles are satellites of the main peak. Intensity of the main peak falls, intensity of satellites increases with temperature. Redistribution of spectral weight between quasiparticles involving ground and quasiparticles involving excited single-hole eigenstates causes the changes in width and shape of the resulting spectral function peak. Splitting of the LHB on the number of Hubbard polaron subbands and redistribution of spectral weight over these subbands occurs in addition to temperature transformation of $0-0$ quasiparticle band (Fig.~\ref{fig:elstr_Tdep_withEPI}(a),(b)) that has been discussed in the previous Section~\ref{T_dep_without_EPI}. For large EPI ${\lambda _d} = 0.3$ all spectral weight results from the multiphonon excitations. The main effects of temperature growth are broadening of spectral function and damping of its maximal intensity (Fig.~\ref{fig:elstr_Tdep_withEPI}(c)). Value of the FWHM ${\Gamma _{FWHM}}$ grows in $1.3$ times with increasing temperature from $T = 200$~K to $T = 400$~K (inset if Fig.~\ref{fig:elstr_Tdep_withEPI}(c)) whereas in the ARPES ${\Gamma _{FWHM}}$ is doubled~\cite{ShenKM2007}. For the temperatures from $T = 200$~K to $T = 760$~K our calculations show that ${\Gamma _{FWHM}}$ increases more than twice. Thus broadening of spectral function with increasing temperature is qualitatively in agreement to ARPES spectra.

The temperature effect on spectral function depends on momentum value (Fig.~\ref{fig:elstr_Tdep_withEPI}(d)). Shift of the coherent high-intensity peak of the electron spectral function in the valence band has different value and direction at different $k$-points. This shift is caused by reconstruction of quasiparticle band and results from the splitting of $0-0$ band due to hybridization and the weakening of the spin-spin correlations. Direction of the shift of the main peak at the points ${\bf{k}} = \left( {{\pi  \mathord{\left/
 {\vphantom {\pi  2}} \right.
 \kern-\nulldelimiterspace} 2},{\pi  \mathord{\left/
 {\vphantom {\pi  2}} \right.
 \kern-\nulldelimiterspace} 2}} \right)$ and ${\bf{k}} = \left( {0.67\pi ,0} \right)$ is in agreement with the peak shift in the ARPES experiments~\cite{Kim2002} at the same points of momentum space. It is seen that at the points ${\bf{k}} = \left( {\pi ,0} \right)$ and ${\bf{k}} = \left( {\pi ,\pi } \right)$ coherent peak shifts to higher electron energies (lower binding energy in ARPES) (Fig.~\ref{fig:elstr_Tdep_withEPI}(d)). It results in the decreasing of the insulator gap with increasing temperature. This fact is in agreement with data of reflectivity measurements and ${\varepsilon _2}$ spectra in the La$_2$CuO$_4$~\cite{Kastner1998}.

\section{Conclusion \label{Summary}}
The temperature dependence of electronic spectra calculating in the frameworks of p-GTB method is in qualitative agreement with ARPES spectra. Regime of large EPI was considered to reproduce broad peak of spectral function similar to that observed in the ARPES spectra in the undoped cuprates. We have obtained broadening of the peak of the electron spectral function, the reduction of its maximal intensity, shift of the peak and the decreasing dielectric gap with increasing temperature. The FWHM of the spectral function peak at $400$~K is $1.5$ times larger than at $T = 10$~K and $2.6$ times larger at $760$~K. The peak shift is smaller than in ARPES data.

\begin{acknowledgements}
We acknowledge financial support by RFBR (grant 16-02-00098), the research project of Russian Academy of Science 0358-2015-0006 and Government Support of the Leading Scientific Schools of the Russian Federation (NSh-7559.2016.2).
\end{acknowledgements}



\end{document}